# AI-Assisted Geometric Analysis of Cultured Neuronal Networks: Parallels with the Cosmic Web

Wolfgang Kurz[1*] and Danny Baranes[2†]

1 Institute for Measurement Systems and Sensor Technology, TUM School of Computation, Information and Technology, Technical University of Munich, Munich, Germany
2 Department of Molecular Biology, School of Medicine, Ariel University, Ariel, Israel

\* wolfgang.kurz@tum. , † dannyb@ariel.ac.il

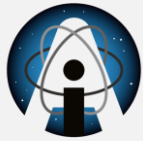

## Abstract

Building on evidence of structural parallels between brain networks and the cosmic web [1], we apply AI-based geometric analysis to cultured neuronal networks. Isolated neurons self-organize into dendritic lattices shaped by reproducible wiring rules. These lattices show non-random features—frequent dendritic convergence, hub nodes, small-world connectivity, and large voids. Synaptic contacts cluster and strengthen at hubs. Strikingly, these properties mirror the cosmic web: dendritic branches resemble cosmic filaments and synapses map to galaxies. Quantitative metrics align across systems, suggesting shared underlying geometric principles. We invite cross-disciplinary collaboration to interrogate and extend these parallels.

## 1 Introduction

**The Cosmic Web**
The cosmic web is a large-scale network of clusters, filaments, sheets, and voids [2]. In ΛCDM, nearly uniform matter collapses into halos (clusters) linked by filaments, with vast voids between themfrontiersin.org. Cosmological simulations (e.g., Millennium) reproduce this filamentary network. Its geometry is quantified via node degree, clustering coefficient, path lengths, and void statistics—measures that capture galaxy clustering and filament interconnections on cosmological scales [3].

**Brain Networks**
The brain's neuronal network is a complex, multiscale graph of cells and connectionsfrontiersin.org. Neurons form circuits/regions—the connectome—and at the cellular level build dense 3D synaptic networks [1]. Neural systems often show small-world organization (high clustering, short paths) [4]; even dissociated in-vitro networks evolve small-world topologies [5]. The detailed geometry of dendritic arborization and its role in patterning connectivity remains under study.

**Cross-Domain Comparisons**
Both systems appear as filament–node "webs" with similar degree distributions, clustering, and path lengths in brain and cosmic graphs [1], suggesting potential common network principles despite disparate scales and physics. Here we use neuronal cultures as a tractable testbed, applying cosmology-inspired graph analysis to dendritic architectures.

**Neuronal Cultures as Models**
Dissociated rodent neurons on glass generate dendritic/axonal arbors imaged by MAP2, neurofilament, and synaptophysin labeling, enabling direct measurement of neurite geometry and synapse locations. Cultures inherently tend toward small-world organization [5, 6], making them suitable for uncovering wiring rules.

**Wiring Rules in Culture**
Imaging reveals non-random motifs: growth toward neurite–neurite contacts forming stable multi-neurite intersections (MNIs) with clustered synaptic markers [7]; coordinated dendritic convergence yielding "economical small-world" hubs [6]; and regulated crossing angles that bias axonal traversal and drive synaptic clustering near intersections [8]. Together, convergence, bundling, and preferred intersection geometry shape where synapses form and the overall lattice.

**AI in Geometric Analysis**
AI supports segmentation, tracing, and graph construction from fluorescence images; e.g., software that maps intersections to topological graphs (Pinchas et al.). We adopt a similar pipeline to extract branch graphs and compare them with cosmic-web analyses. In parallel, ML efforts in cosmology (e.g., CAMELS) learn filamentary structure from simulated universesimonsfoundation.org, enabling a unified framework for cross-scale geometry.

# Methods

**Neuron Culture Data:** Primary hippocampal and cortical neurons were dissociated from postnatal rats and plated on poly-D-lysine/laminin substrates. Cultures were maintained up to 2–3 weeks. Cells were fixed and immunolabeled: anti-MAP2 for dendrites, anti-neurofilament (NFM) for axons, and anti-synaptophysin for synapses and observed through high-resolution fluorescence microscope. Image analysis was performed using the ImageJ software.

**AI Analysis Pipeline:** A deep-learning model segmented neuronal processes (U-Net style CNN trained on labeled data). The segmented images were skeletonized and a graph representation was constructed: nodes were defined at neurite intersection points or branch junctions, and edges corresponded to dendritic segments between nodes. Synapse centroids (from synaptophysin signal) were mapped onto the graph to compute synaptic density around nodes vs. along edges.

# Results

**Dendritic Lattice Geometry**
Cultured neurons self-organize into a dendritic lattice with distinctive geometric features. The wiring is highly non-random: dendrites frequently converge at common junctions, forming hubs of connectivity. These convergence points involve branches from multiple neurons (multi-neurite intersections). Between these hubs, regions of low branch density appear as *voids*. In the network graph, most nodes have low degree (2–3), but a few hubs have many incoming dendrites, yielding a heavy-tailed degree distribution. Figure 1 illustrates a typical dendritic lattice: parallel branch bundles span the field, intersect at hubs (highlighted), and leave large empty areas in between. These results confirm that dendritic arborization is governed by consistent wiring rules: convergence, bundling, and looped paths dominate the lattice topology.

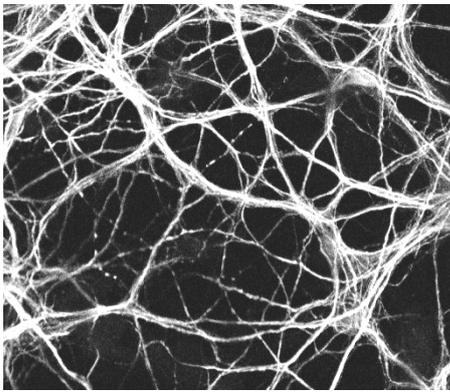

**Figure 1: Dendritic lattice in a cultured neuronal network.** The image shows neuronal processes in a 2-week-old culture. Convergence hubs where multiple dendrites meet, are abundant, generating large black sparse regions (voids). Image scale: 100μm X 100μm.

**Synapse Topography**
The lattice geometry directly shapes synapse placement. We found a pronounced accumulation of synapses at dendritic hubs (see an example in figure 2). Analysis showed that the integrated synaptic signal within small regions centered on convergence nodes is significantly higher than in regions along single branches. This enrichment mirrors previous reports [6, 7]: axonal terminals at intersections exhibited higher synaptophysin content than elsewhere. In our data, hubs (nodes of degree ≥4) often host dense synaptic clusters, whereas low-degree branches have sparse synapses. Thus, dendritic hubs serve as synapse "magnets". This pattern is analogous to cosmic clustering: galaxies concentrate at filament intersections while filaments themselves have fewer galaxies. In other words, neuronal and cosmic networks both display *hub-and-spoke* synaptic topology, with activity or luminosity enhanced at nodes. Our results confirm that dendritic lattice geometry patterns the spatial topology of synaptic connectivity.

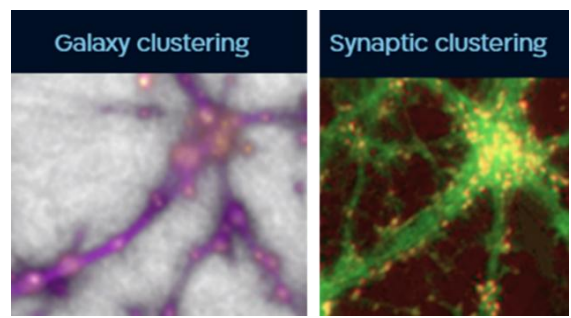

**Figure 2. Morphological similarities between a simulated cosmic web and a neuronal culture.** Left: Simulated cosmic web (purple) with filaments connecting dense galaxy-cluster nodes (pink–brown dots) and enclosing large void regions (gray). *Image credit: NASA, ESA, and J. Burchett and O. Elek (UC Santa Cruz). Licensed under CC BY 4.0.* Right: Cultured neuronal network showing a dendritic hub (green) enriched with synaptic puncta (yellow).

**Comparison with Cosmic Web Patterns**
The parallels between dendritic lattices and the cosmic web are striking. In the cosmic web, galaxy clusters occur where multiple filaments intersect, separated by large voids [1-3]. Similarly, dendritic branches (edges) converge into hub nodes, with void-like spaces between. Figure 2 shows a simulated cosmic web: filaments (light) connect dense clusters (bright), with cosmic voids in between. Both networks exhibit skewed degree distributions (few high-degree hubs, many low-degree nodes) and elevated clustering. For example, the topological clustering coefficient is of order 0.4–0.6 in both cases.

Thus, metrics like degree, clustering, and void fraction fall in comparable ranges (order-of-magnitude) for neurons and galaxies [1]. While the underlying physics differ, these quantitative similarities support the qualitative match observed: dendritic filament ≈ cosmic filament, synapse cluster ≈ galaxy cluster.

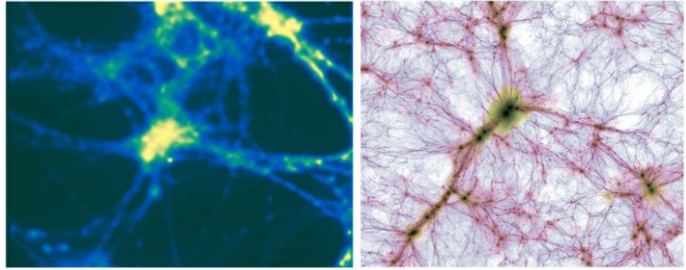

**Figure 3. Enrichment at convergence sites in both networks.** (Left) Convergence sites of neuronal processes (blue) show increased synaptic activity (yellow). Width = 60mm. (Right) Cropped image from thin slice through the cosmic large-scale structure from the IllustrisTNG TNG300 simulation. Brightness encodes projected baryonic mass density; hue encodes mean projected gas temperature. Width ≈ 0.4 billion light-years. Credit: TNG Simulations / TNG Collaboration (IllustrisTNG, TNG300).

## Discussion

The present study highlights shared geometric features of neural and cosmic networks. Both systems form spatial graphs of linear elements (dendrites or filaments) that intersect into hubs and leave voids between. In both, hub nodes concentrate activity – synapses in neurons, galaxy luminosity in clusters – and enable efficient connectivity. These commonalities suggest that universal principles (e.g. optimization of wiring and connectivity) may operate in very different contexts [1].

However, important differences limit this analogy. Neuronal networks are active, adaptive biological systems, whereas the cosmic web is a relatively static gravitational structure. The formation laws are fundamentally different (cell guidance cues vs. gravity). The scales differ by ~20 orders of magnitude, and neuronal networks in vitro are essentially two-dimensional (culture) versus the 3D universe. These differences mean that our comparison is suggestive rather than literal: the shared geometry likely reflects common graph dynamics (e.g. growth toward connection points) rather than a direct physical link. Vazza & Feletti similarly caution that visual likeness does not imply identical mechanisms [1].

Finally, AI tools were instrumental in this analysis. Automated image segmentation and graph extraction allowed us to quantify subtle patterns in large datasets. In cosmology, projects like CAMELS demonstrate how ML can decode complex structures from simulations [9]. By analogy, our deep-learning pipeline efficiently identified dendritic intersections and computed network metrics that would be time-consuming by hand. In future work, enhanced AI models (e.g. graph neural networks) could further bridge neural and cosmic datasets, potentially uncovering deeper organizing laws.

## Conclusion

Cultured neuronal networks form intricate dendritic lattices defined by reproducible wiring rules. These lattices exhibit hubs and voids and non-random branch convergence. We found that synapses concentrate at dendritic convergence points. Strikingly, these features parallel the cosmic web's filament-node structure. While the two systems differ fundamentally, the shared geometry raises open questions: might there be universal graph-growth principles at work? Can insights from one domain inform the other? Addressing these questions requires interdisciplinary effort. We encourage collaborations between neuroscientists and cosmologists to explore these intriguing parallels and underlying mechanisms.

## Acknowledgements

The authors would like to thank Dr. FFF for assistance with the preparation of our conference presentation.

**Funding information** This work was not funded